%% file: tr.tex
\documentclass[12pt]{article}
\pdfoutput=1
\usepackage{fullpage}
\usepackage{times}
\usepackage{url}
\usepackage{graphicx}
\usepackage{color}
\usepackage{amsmath}

\input{makros.tex}

\newcommand{\Id}[1]{\ensuremath{\mathit{#1}}}
\newcommand{\emptyEntry}{\bot}
\newcommand{\tombstone}{\dag}
\newcommand{\hmin}{\check{h}}

\newcommand{\nopetaedge}[1]{}

\newcommand{\mytitle}{Hashing with Linear Probing and Referential Integrity}
\definecolor{mygray}{rgb}{0.5,0.5,0.5}
\definecolor{myred}{rgb}{1,0.5,0.5}
\definecolor{myblue}{rgb}{0.5,0.5,1}
\definecolor{mygreen}{rgb}{0.4,0.8,0.4}

\begin{document} 


\title{\mytitle}
\author{Peter Sanders\\
Karlsruhe Institute of Technology (KIT), 76128 Karlsruhe, Germany\\ 
\url{sanders@kit.edu}}
\date{\today}
\pagestyle{plain}

\maketitle
\begin{abstract}
  We describe a variant of linear probing hash tables that never moves
  elements and thus supports referential integrity, i.e., pointers to
  elements remain valid while this element is in the hash table. This is
  achieved by the folklore method of marking some table entries as
  formerly occupied (tombstones). The innovation is that the number of
  tombstones is minimized. Experiments indicate that this allows an
  unbounded number of operations with bounded overhead compared to
  linear probing without tombstones (and without referential integrity).
\end{abstract}
\section{Introduction}

Hash tables are among most fundamental and widely used data
structures.  Refer to \cite{MehSan08} for examples and more detailed
discussion of the basic techniques. While there is a plethora of hash
table data structures, hashing with linear probing is the most
efficient one in many practical situations.  This is due to its
simplicity, cache efficiency, absence of overhead for internally used
pointers, and because only a single hash function evaluation is needed
for a search or insert operation.

However, deletion is problematic for linear probing.  There are two
main known approaches. Usually the preferred one is to rearrange
elements so that the main data structure invariants are maintained
\cite{Knu98,MehSan08}. However this destroys referential integrity --
pointers to moved elements are no longer valid\footnote{As an
  illustrative example where referential integrity is relevant,
  consider a LRU (least recently used) cache. An efficient folklore
  implementation consists of a hash table with one entry for each cached object
  and a doubly linked list that remembers how recently elements
  have been accessed (e.g., \cite{RobDev90}). Directly storing list items in the hash table
  requires referential integrity.}. A folklore solution is to simply
mark table entries previously occupied by deleted elements using a
special \emph{tombstone} value. This has the drawback that elements
are never ever freed and thus search costs grow until, eventually, the
entire table has to be searched during unsuccessful searches. Addressing this
problem by reorganizing the table from time to time also
destroys referential integrity.

The idea behind the present paper is to use tombstones but to remove
those that are not needed to maintain the data structure
invariants. Experiments (Section~\ref{s:experiments}) indicate that
this is sufficient to keep search costs bounded as long as the table
is not too full.

\section{Preliminaries}

Suppose we want to store $n$ elements in a table $t[0..m-1]$ of
elements where $m>n$ is the table size.  Let
$h:\Id{Element}\rightarrow 0..m-1$ denote the hash function.  We also
consider special elements $\emptyEntry$ for empty table entries and
$\tombstone$ for tombstones.  The invariant governing the
implementation of linear probing is that if element $e$ is stored at
table entry $t[i]$ then 
the entries cyclically between $t[h(e)]$ and $t[i]$ are nonempty.
Searching an element $e$
then amounts to scan $t$ cyclically starting at $t[h(e)]$.
``Cyclically'' here means the the search wraps around when the end of
the table is reached.  This process stops if either $e$ is found or an
empty table entry is found. In the latter case, the invariant
guarantees that $e$ is not in the table.  Inserting an element $e$
works similarly to searching.  If an elements with the same key
exists, depending on the desired semantics of insertion, nothing is
done or the element is updated. Inserting a new element overwrites
the 
first empty table entry or tombstone at $t[h(e)]$ or cyclically to the right.

Several implementations for deleting an element $e$ are
possible. Two extremes are simply replacing it by a tombstone or
rearranging the elements so as to avoid tombstones altogether
\cite{Knu98,MehSan08}.  

\section{Linear Probing with Referential Integrity}

Our variant of linear probing is based on two principles:
\begin{enumerate}
\item Never move elements. This entails referential integrity.
\item Use tombstones only when necessary. This means, we maintain the
  invariant that when $t[i]=\tombstone$ then there is an element $e$
  in the table such that $h(e)$ is $i$ or cyclically to the left yet
  $e$ is stored cyclically to the right of $i$.
\end{enumerate}

\begin{figure}
  \begin{code}
    \Procedure delete$(e)$\+\\
      \For $i\Is h(e)$ \While $t[i]\neq e$ \Do\+\RRem{search for $e$}\\
        \If $t[i]=\emptyEntry$ \Then \Return\RRem{$e$ is not in $t$}\-\\
      $t[i]\Is\tombstone$\RRem{tombstone -- may be removed later}\\
      $\hmin\Is m$\RRem{initialize smallest hash function value encountered}\\
      \For $j\Is i+1$ \While $t[j]\neq\emptyEntry$\RRem{scan to the right}\+\\
        \If $t[j]\neq\tombstone$ \Then
          \If $h(t[j])<\hmin$ \Then $\hmin\Is h(t[j])$\RRem{update smallest hash function value}\-\\
      \For $k\Is i$ \Downto $h(e)$ \Do\+\RRem{scan to the left}\\
        \If $t[j]=\tombstone$ \Then
          \If $\hmin>k$ \Then $t[j]\Is\emptyEntry$\RRem{remove tombstone}\\
        \Else
          \If $h(t[j])<\hmin$ \Then $\hmin\Is h(t[j])$\RRem{update smallest hash function value}     
  \end{code}
\caption{\label{fig:delete}Pseudocode for deletion in linear probing with referential integrity. To keep the code simple, we describe a variant without wrap-around, i.e., $t$ is allocated sufficiently large such that overflowing elements always find free entries there; see also \cite{MehSan08}.}
\end{figure}

The main issue is now how to maintain this invariant efficiently.
Searches, updates and insertions do not affect the invariant and can
be implemented as before.

Deletion has two aspects relevant for the invariant.  On the one hand,
a deleted element $e$ previously stored at position $i$ has to become
a tombstone if there are elements hashed to $i$ or cyclically to the left
but stored cyclically to the right of $i$. On the other hand,
tombstones cyclically between $h(e)$ and $i$ may become unnecessary
because they were only needed to be able to find $e$.  Both aspects
can be handled in a uniform way.  Let $j$ denote the position of the
first empty cell cyclically to the right of $i$.
Initially, $t[i]$ is replaced by a tombstone.
We then scan
$t[i+1..j-1]$ and compute the hash function value $\hmin$ occurring there that is
cyclically farthest to the left. Then we scan $t[h(e)..i]$ from right to left.
During this scan we update $\hmin$. Moreover, when encountering a tombstone at position $k$,
it is replaced by $\emptyEntry$ if $\hmin$ is cyclically to the right of $k$.
Figure~\ref{fig:delete} gives pseudocode for the deletion operation.

For correctness, first note that only in the range $h(e)..i$ can
tombstones become unnecessary because only $e$ is deleted and other
positions are irrelevant for keeping $e$ searchable. When a tombstone
at position $k$ is removed, this is save since we know that no element
stored cyclically to the right of $k$ is hashed to position $k$ or
cyclically to the left. On the other hand, when a tombstone is
kept, this is necessary since we encountered an element cyclically to the
right of $k$ that is hashed to $k$ or cyclically to the left.

\section{Experiments}\label{s:experiments}
\begin{figure}[htb]
  \begin{center}
    \includegraphics[scale=1.1,trim=0mm 5mm 30mm 0mm]{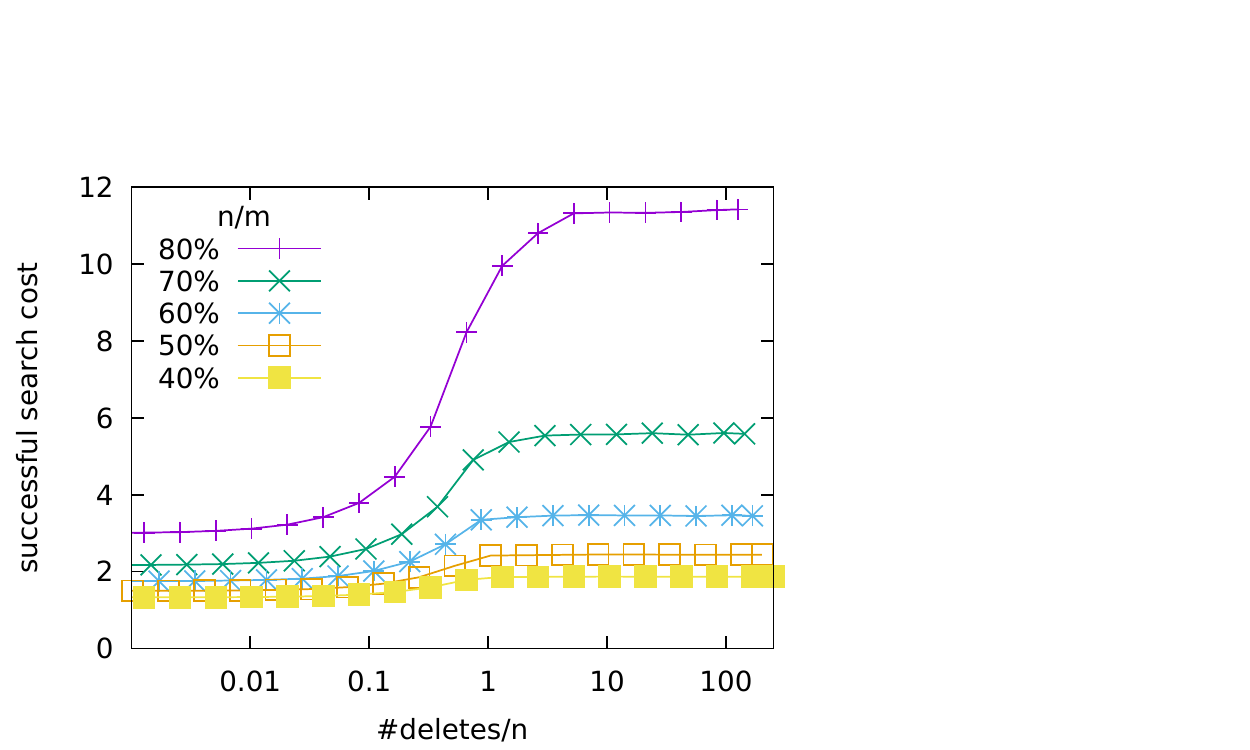}\\
    \vspace*{-10mm}
    \includegraphics[scale=1.1,trim=0mm 5mm 30mm 0mm]{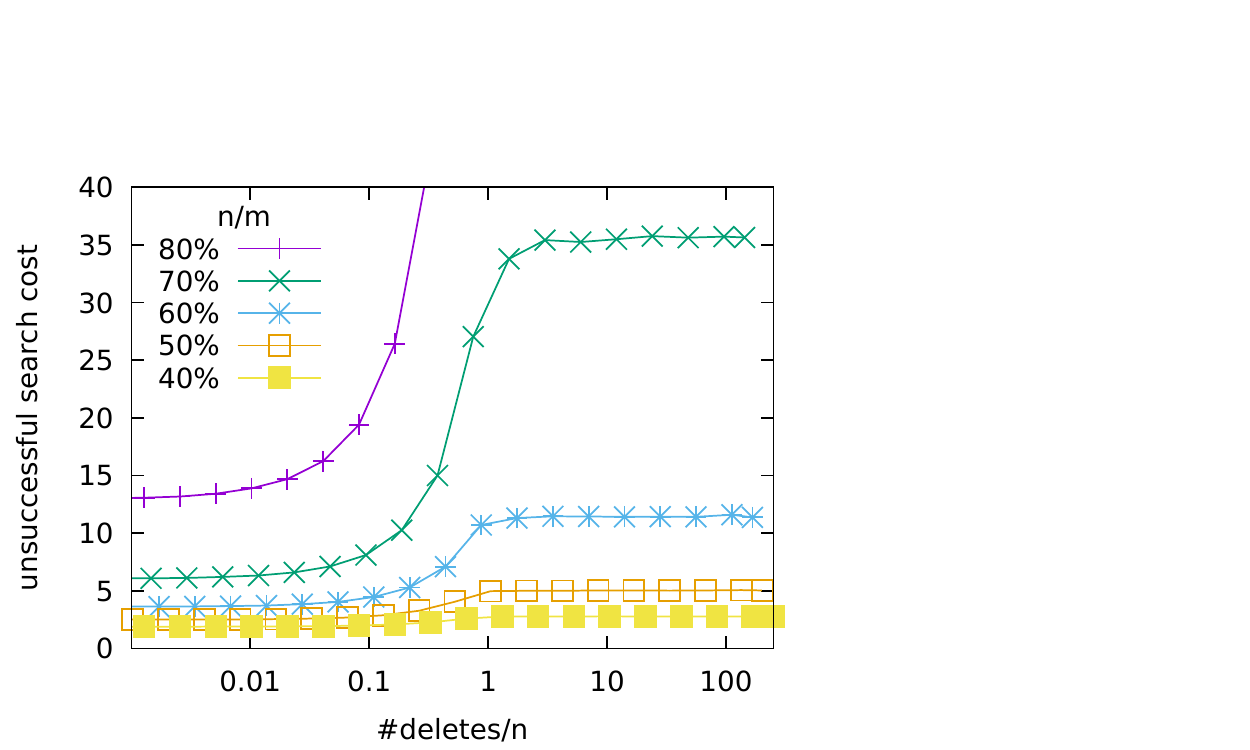}
  \end{center}
  \caption{\label{fig:unsuccessful}Average number of table entries accessed during successful search (top) and unsuccessful search (bottom) as a function of the number of deletions performed ($m=10^6$). For $n/m=80\%$, the unsuccessful access cost converges to around 210.}
\end{figure}
\begin{figure}[htb]
  \begin{center}
    \includegraphics[scale=1.1,trim=0mm 5mm 30mm 0mm]{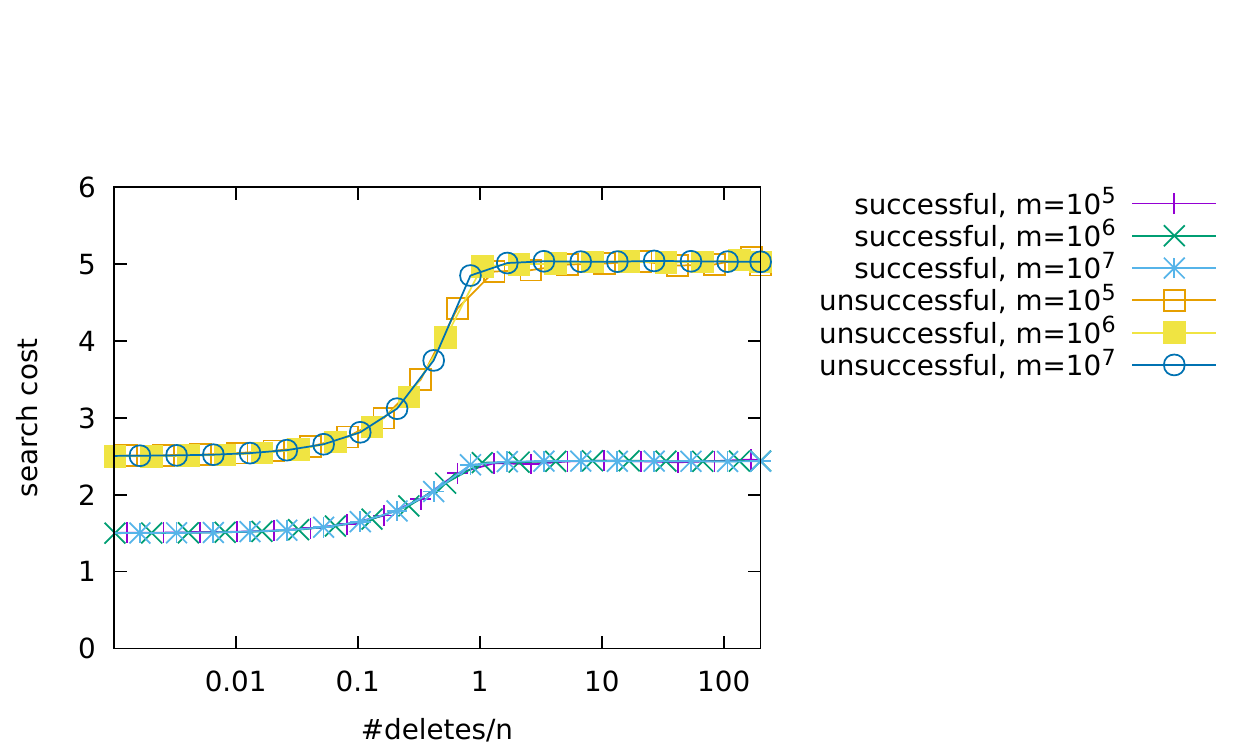}
  \end{center}
  \caption{\label{fig:scaling}Average number of table entries accessed during (un)successful search as a function of the number of deletions performed. We have $n/m=50\%$ in all these experiments and vary $m$.}
\end{figure}

We have performed a number of simple experiments in order to test the
hypothesis that search times remain bounded regardless of the number
of operations performed on the hash table (in contrast to naively
always putting tombstones). We also wanted to get an idea about the
overheads compared to avoiding tombstones altogether. In the
experiments below, we begin by inserting $n$ elements. Then we
alternate between removing the least recently inserted element and
inserting a new element.\footnote{The source code is available at \url{http://algo2.iti.kit.edu/sanders/programs/hash/}. To avoid conceivable issues due to weak hash functions, we use tabulated values of a high quality pseudo random number generator \cite{MatNis98}.}

Figure~\ref{fig:unsuccessful}-top shows the average number of table entries
accessed for successful search. The cost increase is moderate up to $n/m\approx 60\%$.
Figure~\ref{fig:unsuccessful}-bottom gives the corresponding numbers for unsuccessful searches.
Consistently with linear probing in general, these costs are considerably higher.
Indeed, they get very high for load factor $80\%$.
Figure~\ref{fig:scaling} indicates that the average cost bounds are independent of the overall table size.

We have also performed similar experiments where the deleted element is chosen randomly. The results are qualitatively similar -- bounded access times independent of $n$ for fixed $n/m$.
Quantitatively, successful search times are significantly smaller whereas unsuccessful search times are slightly larger.

\section{Conclusions}
We have presented a variant of linear probing hash tables that
preserve referential integrity under deletions. First experiments
indicate that this yields good performance as long as the table is not
too full. Further experiments could give additional evidence. Even
better would be a theoretical analysis.
Note that analyzing deletions in open addressing hashing is not easy in general. Perhaps one can adapt the approach of Mitzenmacher \cite{Mitz16} used for Robin Hood hashing and assuming that
deletions pick random elements and that insertions are new elements.

\bibliographystyle{plain}
\bibliography{diss}

\end{document}

%% file: makros.tex
\usepackage{amsfonts}

\newcommand{\realrange}[2]{\left[#1, #2\right]}

\newcommand{\unitrange}[2]{\realrange{0}{1}}

\newcommand{\llabel}[1]{\label{\labelprefix:#1}}
\newcommand{\labelprefix}{} 

\newcommand{\discussionsize}{\small}

\newenvironment{code}{\noindent
\begin{tabbing}%
\hspace{2em}\=\hspace{2em}\=\hspace{2em}\=\hspace{2em}\=\hspace{2em}\=%
\hspace{2em}\=\hspace{2em}\=\hspace{2em}\=\hspace{2em}\=\hspace{2em}\=%
\kill}{\end{tabbing}}

\newcommand{\labelcommand}{}
\newcommand{\captiontext}{}
\newsavebox{\codeparam}
\newcounter{lineNumber}
\newenvironment{disscodepos}[3]{%
\renewcommand{\labelcommand}{#2}%
\renewcommand{\captiontext}{#3}%
\sbox{\codeparam}{\parbox{\textwidth}{#3}}%
\begin{figure}[#1]\begin{center}\begin{code}\setcounter{lineNumber}{1}}{%
\end{code}\end{center}\caption{\llabel{\labelcommand}\captiontext}\end{figure}}

{\end{disscodepos}}

\newcommand{\Procedure}{{\bf Procedure\ }}

\newcommand{\While}    {{\bf while\ }}

\newcommand{\Do}       {{\bf do\ }}

\newcommand{\For}      {{\bf for\ }}

\newcommand{\Downto}       {{\bf downto\ }}
\newcommand{\If}       {{\bf if\ }}
\newcommand{\Is}       {:=}

\newcommand{\Then}     {{\bf then\ }}
\newcommand{\Else}     {{\bf else\ }}

\newcommand{\Return}   {{\bf return\ }}

\newcommand{\RRem}[1]   {\`{\bf --\hspace{0.5mm}--~}{\rm#1}}

\newdimen\endofsize\endofsize=0.5em
\def\endofbeweis{~\quad\hglue\hsize minus\hsize
                 \hbox{\vrule height \endofsize width
\endofsize}\par}
